\documentclass[preprint,aps,showpacs,preprintnumbers,amsmath,amssymb]{revtex4}
\date{\today}

\usepackage{graphicx}

\newcommand{\Integer}{\mathbb{Z}}

\begin{document}



\title{Saturation of the production of quantum entanglement between 
weakly coupled mapping systems in strongly chaotic region}

\author{Atushi Tanaka}
\email{tanaka@phys.metro-u.ac.jp}
\affiliation{%
Department of Physics, Tokyo Metropolitan University,
Minami-Osawa, Hachioji, Tokyo 192-0397, Japan 
}%

\author{Hiroshi Fujisaki}
\email{fujisaki@ims.ac.jp}
\affiliation{%
Department of Theoretical Studies, Institute for Molecular Science,
Myodaiji, Okazaki 444-8585, Japan
}%

\author{Takayuki Miyadera}
\email{miyadera@is.noda.tus.ac.jp}
\affiliation{%
Department of Information Sciences, 
Tokyo University of Science, Noda City, 
Chiba 278-8510, Japan
}%

\begin{abstract}
 The production of quantum entanglement between weakly coupled mapping 
 systems, whose classical counterparts are both strongly chaotic, is
 investigated. 
 In the weak coupling regime, it is shown that time correlation functions
 of the unperturbed systems determine the entanglement production.
 In particular, we elucidate that the increment of nonlinear parameter
 of coupled kicked tops does not accelerate the entanglement production
 in the strongly chaotic region. 
 An approach to the dynamical inhibition of entanglement is suggested. 
\end{abstract}

\pacs{05.45.Mt,03.65.Ud,05.70.Ln,03.67.-a}

\maketitle


In a quantum composite system, even if the subsystems are remotely
separated and the whole system is in a pure state, the subsystems
generically have a nonclassical correlation~\cite{EPR-Bell}. This
striking phenomena is called quantum entanglement~%
\cite{Schroedinger:PCPS-31-555}, which is utilized not only to achieve
the procedures that have no classical analogs (e.g. quantum information
processing~\cite{QIP}), but also to realize the ``classical world'' in
which quantum interference phenomena are ``decohered'' as a result of
quantum dynamics~\cite{Decoherence}.
Even when there is no quantum entanglement between subsystems, a weak
interaction between the subsystems generally produces the quantum
entanglement during unitary time evolutions~\cite{Kubler:AP-76-1973}.  
This is an important dynamical origin of
decoherence~\cite{Decoherence}.

Through a number of numerical experiments, it is known that the
productions of entanglements induced by quantum dynamics heavily depend
on the qualitative nature of the corresponding classical dynamics,
namely, regular or chaotic~%
\cite{Adachi:1992,Tanaka:JPA-29-5475,Sakagami:PTP-95-703,Miller:PRE-60-1542},
as is easily expected from the studies of 
``quantum chaos''~\cite{Gutzwiller:CCQM-1990}. 
On one hand, in classically regular systems, the confinement of
phase-space dynamics in a narrow region enclosed by KAM tori make it
difficult to produce strong entanglements in
general~\cite{RegularException}.  
On the other hand, the absence of such dynamical barriers in classically
chaotic systems promotes the production of entanglement.
Although there are quantum effects on the phase-space dynamics,
e.g. tunnelings and localizations~\cite{Localizations},
in both regular and chaotic systems, it is confirmed that the scenario
above qualitatively holds~%
\cite{Adachi:1992,Tanaka:JPA-29-5475,Sakagami:PTP-95-703,Miller:PRE-60-1542}.

This motivates the next question: In the chaotic region, does
stronger chaos enhance the production of entanglement? Looking for an
analogy of a study on quantum open systems~\cite{Zurek:PRL-72-2508}, 
Miller and Sarkar obtained a numerical result that suggests the
linear instability of classical dynamics enhances the production of
entanglement~\cite{Miller:PRE-60-1542}. Their numerical experiment
however concerns only in the {\em weakly} chaotic region where chaotic
seas and tori coexist. 
The complexity of phase-space dynamics is the source of the difficulty 
to obtain a theoretical explanation for the Miller and Sarkar's result.

Our aim is to provide a theoretical argument of entanglement
production in weakly coupled chaotic systems. In contrast to the
Miller and Sarkar's work, we focus on the {\em strongly} chaotic
region where the effect of tori is small, to facilitate to obtain a
theoretical explanation. 
Starting from separable pure states, we examine the productions of
quantum entanglement due to unitary time evolutions.
The entanglement production processes are slow due to the weak
coupling. Furthermore, the recurrence time of classically chaotic
systems is relatively long. Hence, the entanglement production
processes are nearly stationary processes, at least, in a short time
period.
This enables us to introduce an entanglement
production rate. We investigate how the entanglement production rate 
depends on the nonlinear parameter below.


Our numerical experiments employ coupled
kicked tops~\cite{Miller:PRE-60-1542}. 
Firstly, we introduce their constituent, a kicked
top~\cite{Haake:ZPB-65-381}, which is described by the Hamiltonian
\mbox{$\hat{H}_k \equiv \pi\hat{J}_y /2
+ \Delta(t)\; k \hat{J}_z^2 / (2j) $}, 
where $\hat{J}_i$ is the $i$-th component of the angular momentum
operator of the top, 
$j$ is the magnitude of the angular momentum, $k$ is a nonlinear
parameter, and $\Delta(t)\equiv\sum_{n\in\Integer}\delta(t-n)$ is 
a ``periodic delta function''.
Secondly, we employ the following Hamiltonian to describe the whole
system that is composed of two kicked tops:
\begin{equation}
\hat{H} \equiv \hat{H}_{k_1}\otimes\hat{1}
+ \hat{1}\otimes\hat{H}_{k_2} + \epsilon\;\hat{V}\Delta(t)
\end{equation}
where $\epsilon$ is a coupling constant, 
$\hat{V} \equiv \hat{J}_{1z}\hat{J}_{2z}/j$ is the interaction 
Hamiltonian, 
and $J_{iz}$ is the $J_z$ of the $i$-th top. 
We report the case that the magnitudes of the angular momenta of the two 
subsystems are the same value, $j$.

Since we focus on the case where the total system is in a pure
state, our choice of a measure of quantum entanglement between the two
subsystems, is the von Neumann entropy $S_{\rm vN}$ of the first
subsystem~\cite{EntanglementMeasure}.
Note that the von Neumann entropy of the second subsystems is equal to
the one of the first subsystem, when the total system is in a pure
state. At the same time, we employ the linear entropy 
$S_{\rm lin}$ instead of $S_{\rm vN}$, to facilitate theoretical
arguments. In our numerical experiments, these entropies behave
qualitatively similarly. 

\begin{figure}
  \includegraphics[width=8.5cm]{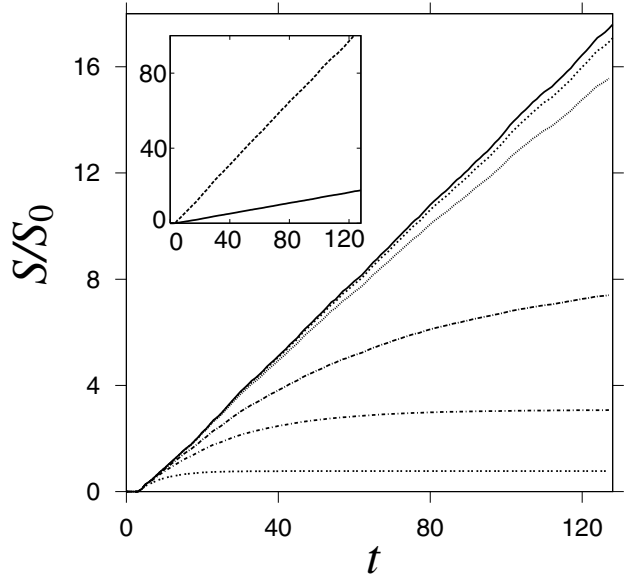}
  \caption{\label{fig:evolS} Time evolutions of quantum entanglement,
    measured by the entropies of a subsystem.
    The entropies are scaled by $S_0 = 2\epsilon^2 j^2$
    (cf. eq.~(\ref{eq:miyafor})). 
%
%
    From bottom to top, we depict $S_{\rm lin}$ with
    $\epsilon = 10^{-2}$, $5\times 10^{-3}$, $3\times 10^{-3}$, 
    $10^{-3}$, $5\times 10^{-4}$, $10^{-4}$. An estimation
    $S_{\rm li n}^{\rm PT}$ by our perturbation
    formula~(\ref{eq:miyafor}) is degenerate with the case 
    $\epsilon = 10^{-4}$.
    In the inset, the solid line and the dotted line correspond to 
    $S_{\rm lin}$ and $S_{\rm vN}$, respectively, at $\epsilon = 10^{-4}$.
    The value of nonlinear parameters are $k_1 = k_2 = 7.0$, which means
    that the corresponding classical tops are strongly
    chaotic~\cite{Haake:ZPB-65-381}.  
    The magnitude of the angular momenta is chosen to be large $j = 80$,
    in order to investigate the semiclassical regime. The center of the
    initial wave packet, which is a product of spin-coherent
    states~\cite{KlauderSkagestram:CS-1985}, is 
    $(\theta_1, \phi_1, \theta_2, \phi_2) = (0.89, 0.63, 0.89, 0.63)$.}
\end{figure}

\begin{figure}
  \includegraphics[width=8.5cm]{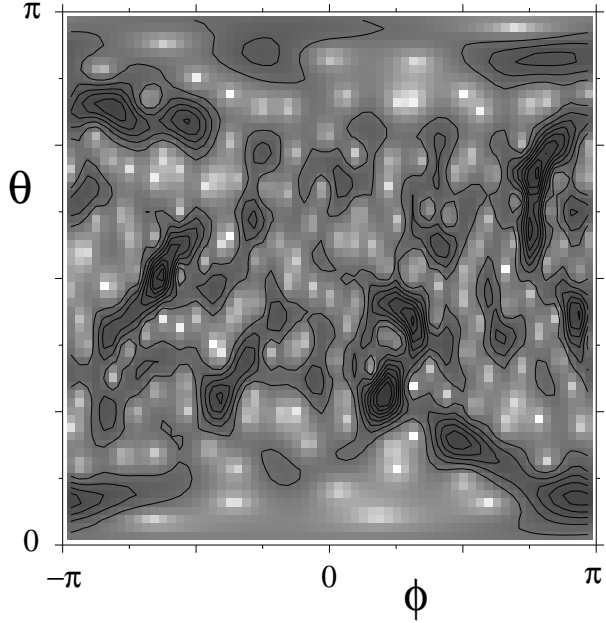}
  \caption{\label{fig:husimi}%
    The Husimi function~\cite{Hushimi} of the first subsystem at 
    $t=15$, during the stationary entanglement production region. 
    The contour and density plots are in normal and logarithmic scales,
    respectively. Note that this region is beyond the Ehrenfest 
    time~\cite{Berry:JPA-12-625}. We choose $\epsilon = 10^{-4}$.
    Other parameters are the same as in Fig.~\ref{fig:evolS}. 
  } 
\end{figure}

We examine the productions of quantum entanglements, by observing
$S_{\rm vN}$ and $S_{\rm lin}$, during unitary time evolutions whose
initial states are product (i.e. separable) states.
A typical result in the strongly chaotic region is shown in
Fig.~\ref{fig:evolS}:
When the coupling constant $\epsilon$ is small,
there is a $t$-linear entanglement production region, which is wide
enough to call it ``stationary'' entanglement production.
Note that during the stationary entanglement production, the state
vectors of the subsystems are spread over the phase space of the
subsystems (Fig.~\ref{fig:husimi}). In other words, the phase space 
distribution of each subsystem is nearly in ``equilibrium''.
We accordingly expect that the each subsystem plays a role of a 
chaotic ``heat bath'' of its partner~%
\cite{Adachi:PRL-61-659}.
The $t$-linear entanglement production region starts at a time step
$T'$, after a short transient to attain the equilibrium of the phase
space distribution of the subsystems, and ends at a time step $T''$,
until the increment of the entropy reaches its equilibrium
(see Fig.~\ref{fig:evolS}, larger $\epsilon$).

In order to explain the $t$-linear, stationary entanglement production,
we employ a time-dependent perturbation theory, whose small parameter
is a coupling constant $\epsilon$, to evaluate the linear entropy
$S_{\rm lin}(t)$ at $t$-th step. The resultant formula is 
\begin{equation}
 \label{eq:miyafor}
 S_{\rm lin}^{\rm PT} (t) =
 S_0 \sum_{m=1}^{t}\sum_{n=1}^{t} D(m, n)
\end{equation}
where $S_0 \equiv 2 \epsilon^2 j^2$, 
and $D(m, n)$ is a time-correlation function of the uncoupled system. 
Furthermore, since the interaction Hamiltonian $\hat{V}$
takes a bilinear form, 
$D(m, n)$ is decomposed as follows
\begin{equation}
 \label{eq:Ddef}
  D(m, n) \equiv C_1(m, n)\; C_2(m, n)
\end{equation}
where 
$C_i (m, n)\equiv
j^{-2} (\langle \hat{J}_{iz}^m\hat{J}_{iz}^n\rangle 
  -\langle \hat{J}_{iz}^m\rangle\langle \hat{J}_{iz}^n\rangle)$
is a normalized correlation functions of 
$\hat{J}_{iz}^n$, which is evolved by the unperturbed Hamiltonian 
$H|_{\epsilon = 0}$ until $n$-th steps, with an initial condition
$\hat{J}_{iz}^0 = \hat{J}_{iz}$, and the expectation value
$\langle\cdot\rangle$ is respect to the unperturbed system.
%
%
%
%
%
The details to obtain the formula~(\ref{eq:miyafor}) will be shown
elsewhere~\cite{FMT020}.

We remark on the entanglement production
formula~(\ref{eq:miyafor}):
(i) Although we start form the evaluation of the entropy of a
subsystem, the formula~(\ref{eq:miyafor}) is in a symmetric form
with respect to the exchange of the two subsystems. 
This is consistent with the symmetric nature of quantum entanglement
when the whole system is in a pure state; 
(ii) Since our approach does not take into account the effect of the
recurrence, the formula~(\ref{eq:miyafor}) would have qualitatively
different applicability to the classically regular and chaotic
systems. On one hand, for classically regular systems, our theory
would break down in relatively short time period, due to the smallness
of the period of the recurrence. On the other hand, for chaotic
systems, we numerically confirmed that our theory works rather long
time period; 
(iii) Our formula has a similarity with the ones in phenomenological
descriptions of linear irreversible processes~\cite{Kubo1985}, in the
sense that these theories use time correlation functions to describe
relaxation phenomena. This is useful both for discussing
phenomenological arguments and for making a link with a
phenomenological theory and a microscopic theory~\cite{Kubo1985}.

\begin{figure}
   \includegraphics[width=0.9\textwidth]{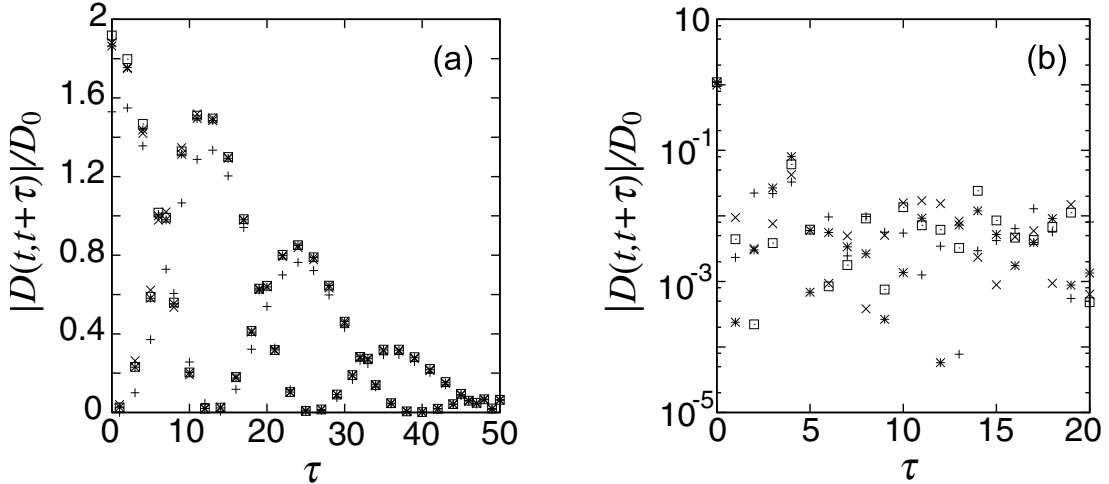}

   \caption{\label{fig:D} The $\tau$ dependence of the correlation function 
   $|D(t + \tau, t)|$ for (a) a regular system ($k_1 = k_2 = 1.0$) and
   (b) a chaotic system ($k_1 = k_2 = 7.0$) with $j = 80$.
   Different symbols correspond to different values of $t$ 
   ($+$: $t=40$, $\times$: $t=50$, $*$: $t=60$, 
   $\square$: $t=70$). Note that (b) employs a normal-log scale.
   } 
  \end{figure}

Before applying the formula~(\ref{eq:miyafor}) numerically, let us
confirm that the time correlation function $D(m,n)$, which is the most
important ingredient of the formula~(\ref{eq:miyafor}), strongly depends
on the dynamics of the classical counterparts (Fig.~\ref{fig:D}).
On one hand, in the regular case, $D(m, n)$ decays slowly with
large oscillations, as the time interval $|m-n|$ become large. On the other
hand, the chaotic dynamics makes the decay of $D(m, n)$ much faster.
Such a rapid convergence of the correlation function, together with
the formula~(\ref{eq:miyafor}), implies the $t$-linear,
stationary entanglement production region (see Fig.~\ref{fig:evolS}). 

The perturbation formula~(\ref{eq:miyafor}) provides
an approximate estimation of the entanglement production rate $\Gamma$
of the $t$-linear, stationary entanglement production region:
\begin{equation}
 \label{eq:GPT}
 \Gamma^{\rm PT}\equiv S_0\; \frac{1}{T''-T'}
  \sum_{m = T'}^{T''}\sum_{n = T'}^{T''} D(m,n)
\end{equation}
where $T'$ and $T''$ are the start and the end of the $t$-linear
entanglement production region, respectively.
In the strongly chaotic region, where the effect of tori is small, it is
possible to give an analytical estimation for 
$\Gamma^{\rm PT}$~(\ref{eq:GPT}). 
Since $D(m, n)$ is the product of the fluctuations of $J_{iz}$, whose
distribution functions become quickly uniform due to the chaotic
dynamics (see Fig.~\ref{fig:husimi}), we assume that $D(m,n)$ decays
exponentially 
\begin{equation}
 \label{eq:Dexp}
 D(m, n) = D_0 \exp(-\gamma |m-n|)
\end{equation}
The prefactor $D_0$ is determined by the magnitude of the
fluctuations of $J_{1z}$ and $J_{2z}$, whose distribution functions are
almost uniform in strongly chaotic systems (see
Fig.~\ref{fig:husimi}). We accordingly assume $D_0 = (1/3)^2$, which is 
independent of $k_1$ and $k_2$.
Furthermore, it is natural to expect that the decay rate $\gamma$ of
$D(m,n)$ increase as the degree of chaos of the classical counterpart
becomes stronger (i.e., as the values of nonlinear parameters $k_1$ and
$k_2$ increase), when the effect of tori is negligibly small,
although we could not obtain a precise nonlinear parameter dependence of 
$\gamma$ due to the fact that the decay is very fast (within a step) in
the strongly chaotic region~(see Fig.~\ref{fig:D} (b)).

\begin{figure}
    \includegraphics[width=6.0cm]{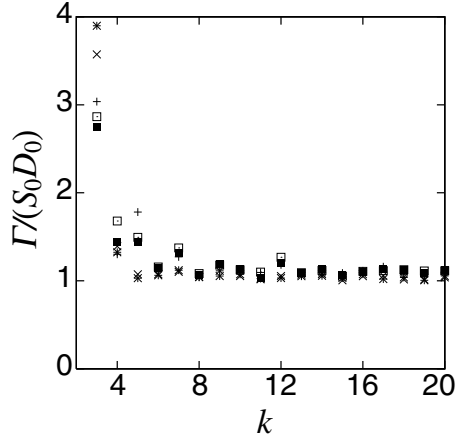}

   \caption{\label{fig:KvsDotS}%
   Dependences of the entanglement production rates $\Gamma$,
   which is measured by linear entropy, on the nonlinear parameter 
   $k = k_1 = k_2$. 
   %
   In order to show typical $k$ dependences, we choose several initial
   conditions (depicted by different marks) that place in the chaotic
   sea. 
   Although the entanglement production heavily depends on the
   initial condition in the weakly chaotic region, the disappearance of
   tori weakens the initial condition dependence in the strongly
   chaotic region.
   %
   %
   Other parameters are the same as in Fig.~\ref{fig:husimi}.
   }
  \end{figure}

As long as the $t$-linear region is wide enough,
i.e. $T''-T'\gg\gamma^{-1}$, the exponential decay 
assumption~(\ref{eq:Dexp}) provides an estimation 
\begin{equation}
 \label{eq:GammaPT}
\Gamma^{\rm PT}\simeq S_0 D_0 \coth(\gamma/2). 
\end{equation}
This provides our main result: in the limit that the
classical counterpart is the strongly chaotic, i.e. in the limit
$\gamma\to\infty$, the perturbation theory~(\ref{eq:miyafor}) predicts
that the entanglement production rate $\Gamma$ converges
to a finite value. Furthermore, the convergence is expected to be fast, 
when $\gamma$ is larger than unity. 
Our main claim is that the effect of the enhancement of entanglement
due to chaotic dynamics saturates in the strongly chaotic region,
in contrast with the weakly chaotic systems 
\cite{Tanaka:JPA-29-5475,Sakagami:PTP-95-703,Miller:PRE-60-1542}. 
This prediction is confirmed by our numerical experiments
(Fig.~\ref{fig:KvsDotS}). 
%
%
We explain the reason why the strongly chaotic systems have smaller
entanglement production rates, in comparison with the weakly chaotic
systems. As the chaos becomes stronger, the
correlation time scale of the interaction Hamiltonian
$\hat{V}$, which is evolved by the unperturbed
Hamiltonian $\hat{H}_0$ in the interaction picture,
becomes shorter. 
Hence, due to the {\em dynamical averaging}, the influence of $\hat{V}$
is effectively reduced. 
Consequently, the entanglement production rate is also reduced.  
In particular, when the chaos is strong enough, the effect of the
perturbation on $\Gamma^{\rm PT}$~(\ref{eq:GammaPT}) comes only from
the ``diagonal'' part $D(n, n)$. This is the origin of the saturation
at large $k$ (Fig.~\ref{fig:KvsDotS}). 

We make a brief remark on our result, in correspondence with 
the existing publications that suggests the entanglement
production rate is proportional to the Lyapunov exponent of the
classical counterpart (i.e., chaos promotes quantum entanglement)%
~\cite{Miller:PRE-60-1542,Zurek:PRL-72-2508}.
%
Firstly, although our model system in the numerical experiments is the
same as Miller and Sarkar's work\cite{Miller:PRE-60-1542}, the result is
qualitatively different. 
%
The difference comes from the different ``strength'' of chaos.
In contrast with the strongly chaotic systems,  the entanglement
productions of the weakly chaotic systems are significantly influenced
by the existence of tori. A crudest explanation is provided by the
perturbation formula~(\ref{eq:GammaPT}):   
During the chaos is not fully developed, a large portion of the phase
space is occupied by tori. This reduces $D_0$, which is the magnitude
of the fluctuation of the interaction $\hat{V}$. Accordingly, as the
chaos become stronger in weakly chaotic region, the development of chaos
increases $D_0$. This results in the increment of the entanglement
production rate.  
In the strongly chaotic region, the growth of $D_0$ saturates due to
the breakdown of tori and $\gamma$ takes a major role in the
parametric variation of $\Gamma^{\rm PT}$~(\ref{eq:GammaPT}).
%
%
%
%
Secondly, we compare our result with Zurek and Paz's
work on open systems~\cite{Zurek:PRL-72-2508}, since the each
subsystem in our numerical experiment acts as ``heat bath'' of its
partner. 
The most important difference comes from the fact that the two works
focus on completely different region, far before (Zurek and Paz) and
far after (ours) the Ehrenfest time (see, Fig.~\ref{fig:husimi}). 
%

We expect that our result on the saturation of entanglement between
weakly coupled systems is generic, in strongly chaotic systems 
with a ``compact'' phase space that allows the
assumption~(\ref{eq:Dexp}). 
However, we note that our result~(\ref{eq:GammaPT}) heavily depends
on the fact that the system is a discrete-time system, i.e. a mapping
system. 

Finally, we discuss an extension of our result to flow systems.
%
%
%
It is straightforward to obtain a perturbative estimation of
entanglement production rate  $\Gamma \propto 1 / \gamma$,
which implies the {\em suppression} of the entanglement productions
in the strongly chaotic limit $\gamma\to\infty$.
%
%
%
This is completely opposite to the case in the weakly chaotic region~%
\cite{Adachi:1992,Tanaka:JPA-29-5475,Sakagami:PTP-95-703,Miller:PRE-60-1542}. 
More thorough investigations on this point will be reported in
subsequent publications. 
We remark that a similar suppression of quantum relaxation due to strong
chaos is reported by Prosen~\cite{Prosen:PRE-65-036208}, in a perturbative
evaluations of fidelity, which is an overlapping integral between
the two states evolved by slightly different Hamiltonians.
As is discussed in~\cite{Prosen:JPA-34-L681}, it is hopeful that
the suppression of quantum relaxations due to strongly chaotic dynamics
will have various application. In particular, our scenario, which
suggests an approach of the {\em dynamical inhibition of entanglement},
will also provide important applications to quantum communications and
computations, which require the protections against
decoherence~\cite{QIP,Prosen:JPA-34-L681}.   

\begin{acknowledgments}
H.~F. thanks T.~Takami and H.~Kamisaka for discussion. 
A.~T. thanks Professor A.~Shudo for useful conversations.
\end{acknowledgments}


\begin{thebibliography}{99}
\bibitem{EPR-Bell}{%
    A. Einstein, B. Podolsky, and N. Rosen,
    Phys. Rev. {\bf 47}, 777 (1935);
    J. S. Bell, Physics {\bf 1}, 195 (1964).}
\bibitem{Schroedinger:PCPS-31-555}{%
    E. Schr\"odinger,
    Proc. Camb. Phil. Soc. {\bf 31}, 555 (1935).}
\bibitem{QIP}{M. A. Nielsen and I. L. Chuang, 
    Quantum Computation and Quantum Information 
    (Cambridge UP, Cambridge, 2000).}
\bibitem{Decoherence}{%
    D. Giulini et al.,
    Decoherence and the Appearance of a Classical
    World in Quantum Theory (Springer, Berlin, 1996);
    W. H. Zurek, e-print quant-ph/0105127 and references
    therein.}
\bibitem{Kubler:AP-76-1973}{O. K\"ubler, and H. D. Zeh, 
    Ann. Phys. {\bf 76}, 405 (1973).}
\bibitem{Adachi:1992}{%
        S.~Adachi, in {\em Proceedings of ISKIT '92},
        edited by I.~Tsuda and K.~Takahashi, 
        (ISIP, Iizuka, 1992), p.~76}
\bibitem{Tanaka:JPA-29-5475}{A.~Tanaka, 
    J. Phys. A: Math. Gen. {\bf 29}, 5475 (1996).}
\bibitem{Sakagami:PTP-95-703}{%
    M. Sakagami, H. Kubotani, and T. Okamura,
    Prog. Theo. Phys. {\bf 95}, 703 (1996);
    K. Furuya, M. C. Nemes, and G. Q. Pellegrino,
    Phys. Rev. Lett. {\bf 80}, 5524 (1998).}
\bibitem{Miller:PRE-60-1542}{P. A. Miller, and S. Sarkar,
    Phys. Rev. E {\bf 60}, 1542 (1999).}
\bibitem{Gutzwiller:CCQM-1990}{M. C. Gutzwiller, 
    Chaos in Classical and Quantum Mechanics
    (Springer-Verlag, New York, 1990).}
\bibitem{RegularException}{%
    Exceptions for classically regular systems are found in
    references~\cite{Tanaka:JPA-29-5475} 
    and~\cite{Angelo:PRE-60-5407}.}
\bibitem{Angelo:PRE-60-5407}{%
    R. M. Angelo et al.,
    Phys. Rev. E {\bf 60}, 5407 (1999).}
\bibitem{Localizations}{%
    G. Casati  et al.,
    Lect. Notes Phys. {\bf 93}, 334 (1979).}
\bibitem{Zurek:PRL-72-2508}{%
    W. H. Zurek, and J. P. Paz, 
    Phys. Rev. Lett. {\bf 72}, 2508 (1994).}
\bibitem{Haake:ZPB-65-381}{F.~Haake, M.~Ku\'s, and R.~Scharf, 
    Z. Phys. B {\bf 65}, 381 (1987).}
\bibitem{EntanglementMeasure}{%
    S. M. Barnett, and S. J. D. Phoenix, Phys. Rev. A {\bf 40},
    2404 (1989).}
\bibitem{KlauderSkagestram:CS-1985}{J. R. Klauder and B.-S. Skagerstam,
    Coherent states (World Scientific, Singapore, 1985).}
\bibitem{Hushimi}{
    K. Husimi, Proc. Phys. Math. Soc. Jpn. {\bf 22}, 264 (1940);
    K. Takahashi, and N. Sait\^o, Phys. Rev. Lett. {\bf 55}, 645
    (1985);
    K. Nakamura et al.,
    Phys. Rev. Lett. {\bf 57}, 5 (1986).}
\bibitem{Berry:JPA-12-625}{M. V. Berry, and N. L. Balazs,
    J. Phys. A {\bf 12}, 625 (1979).}
\bibitem{Adachi:PRL-61-659}{S. Adachi, M. Toda, and K. Ikeda,
    Phys. Rev. Lett. {\bf 61}, 659 (1988).}
\bibitem{FMT020}{H. Fujisaki, T. Miyadera, and A. Tanaka,
    unpublished.}
\bibitem{Kubo1985}{See, e.g., R. Kubo, M. Toda, and N. Hashitsume,
    Statistical Physics II (Springer-Verlag, 1985), \S~3.1.}
\bibitem{Prosen:PRE-65-036208}{%
    T. Prosen, Phys. Rev. E {\bf 65}, 036208 (2002);
    T. Prosen and M. \v{Z}nidari\v{c},
    J. Phys. A: Math. Gen. {\bf 35}, 1455 (2002).}
\bibitem{Prosen:JPA-34-L681}{%
    T. Prosen and M. \v{Z}nidari\v{c},
    J. Phys. A: Math. Gen. {\bf 34}, L681 (2002).}
\end{thebibliography}
\end{document}